\documentstyle[twocolumn,aps,epsf]{revtex}
\begin{document}
\draft
\twocolumn[\hsize\textwidth\columnwidth\hsize\csname @twocolumnfalse\endcsname

%%%%%%%%%%%%%%%%%%%%%%% Title and Authors %%%%%%%%%%%%%%%%%%%%%%%%%%
\title{Stripes Induced by Orbital Ordering in Layered Manganites}

\author{Takashi Hotta$^1$, Adrian Feiguin$^2$, and Elbio Dagotto$^2$}

\address{$^1$Institute for Solid State Physics, University of Tokyo,
5-1-5 Kashiwa-no-ha, Kashiwa, Chiba 277-8581, Japan}

\address{$^2$National High Magnetic Field Laboratory, Florida State
University, Tallahassee, Florida 32306} 

\date{\today}
\maketitle

%%%%%%%%%%%%%%%%%%%%%%%% Abstract %%%%%%%%%%%%%%%%%%%%%%%%%%%%%%%
\begin{abstract}

Spin-charge-orbital ordered structures in doped layered manganites are
investigated using an orbital-degenerate double-exchange model tightly
coupled to Jahn-Teller distortions. 
In the ferromagnetic phase, unexpected diagonal stripes at $x$=$1/m$
($m$=integer) are observed, as in recent experiments.
These stripes are induced by the orbital degree of freedom, which forms
a staggered pattern in the background.
A $\pi$-shift in the orbital order across stripes is identified,
analogous to the $\pi$-shift in spin order across stripes in cuprates.
At $x$=1/4 and 1/3, another non-magnetic
phase with diagonal static charge stripes is stabilized at intermediate
values of the $t_{\rm 2g}$-spins exchange coupling. 

\end{abstract}

%%%%%%%%%%%%%%%%%%%%%%%%%%% Pacs %%%%%%%%%%%%%%%%%%%%%%%%%%%%%%%%

\vskip2pc]
\narrowtext
%%%%%%%%%%%%%%%%%%%%%%% Introduction %%%%%%%%%%%%%%%%%%%%%%%%%%%%

Manganese oxides are currently attracting considerable attention
\cite{review}, due to the complex interplay among spin, charge,
and orbital degrees of freedom, which induces a rich phase diagram
as well as Colossal Magneto-Resistant (CMR) properties.
There are clearly two types of dominant states in these compounds.
For example, in perovskite manganites such as La$_{1-x}$Ca$_x$MnO$_3$,
in the region 0.22$<$$x$$<$0.5 a ferromagnetic (FM) metallic phase
is the ground-state at low temperature. On the other hand,
at $x$$>$0.5, a charge-orbital-spin ordered state is stabilized.
The competition between these two states is at the heart of recent
theories that explain the CMR effect in manganites as arising from 
mixed-phase tendencies\cite{ps}.

This two-phase metal-insulator competition and concomitant large MR
effect occurs also at the Curie temperature $T_{\rm C}$, at densities
where a FM phase exists at low temperature.
Several experiments have clearly revealed the mixed-phase
characteristics of manganites near $T_{\rm C}$ \cite{tunneling}.
While it is natural to assume that one of the competing phases is
FM metallic, the properties of the insulating phase are still unknown.
Recently, considerable progress has been made in this context.
Just above $T_{\rm C}$ evidence for the existence of short-range
{\it stripe-like} charge ordering has been obtained with
neutron diffraction and X-ray scattering studies.
For La$_{1-x}$Ca$_x$MnO$_3$\cite{cubic}, 
``diagonal'' stripes, i.e. charge ordering
(CO) along the $(1,1,0)$-direction\cite{note:vector}, appear
for 0.2$\alt$$x$$\alt$0.3, while for $x$$<$0.2, ``bond'' stripes,
i.e. CO along the (1,0,0) or (0,1,0) direction, have been revealed.
For La$_{2-2x}$Sr$_{1+2x}$Mn$_2$O$_7$\cite{bi-layer},
short-range bond stripes have been detected in the wide range
0.3$\leq$$x$$\leq$0.5.
These results lead to the intriguing possibility that the insulating
phase that contributes to the CMR near $T_{\rm C}$ may also be FM but 
with striped features, a remarkable novel result.
Such a state would be puzzling since stripes in cuprates are associated
with the creation of rivers of holes to avoid having the individual
charges ``fighting'' against the antiferromagnetic (AFM) background. 
Thus in a FM state, where hole movement appears optimal, charge is
naively not expected to form stripes, in contradiction with experiments.

In order to understand this puzzling complex problem, in this Letter
the two-dimensional (2D) double-exchange (DE) model coupled to
Jahn-Teller (JT) phonons is investigated using unbiased computational
techniques.
This model has been already successful in reproducing the
A-type AFM state at $x$=0\cite{hotta1}, the CE-state at
$x$=0.5\cite{hotta2}, and the complex structure of the $x$$>$0.5 regime
\cite{hotta3}, as observed in experiments. However, a state as exotic
as containing stripes in a ferromagnetic background
has not been reported until now.
The present effort focuses on the properties of 2D systems, since 
(i) studies in three-dimensional (3D) cases are technically far more
complex, and
(ii) stripe structures identified experimentally in bilayer and 3D manganites 
are here observed in 2D systems as well.
Our main result is the stabilization of {\it FM states with stripe order}
at $x$=$1/m$ ($m$=integer), a surprising result whose origin lies in the
concomitant orbital order.

Regarding single-layered manganites
$\!{\rm La}_{1-x}\!{\rm Sr}_{1+x}\!{\rm Mn}{\rm O}_4$,
the undoped compound is AFM \cite{kawano}, while at $x$$\sim$0.5
a CE-type AFM CO phase has been identified \cite{murakami}.
However, for 0.0$<$$x$$<$0.5, a complex ``spin glass" behavior
has been experimentally observed \cite{moritomo},
indicating that the 2D ground-state properties are basically unknown.
Our results below also indicate that stripe states  with CE-like AFM
characteristics may exist at $x$=1/4 and 1/3, and they could be important
for the physics of single-layered compounds in non-FM regimes.

The Hamiltonian studied here is
\begin{eqnarray}
  H &=& -\sum_{{\bf ia}\gamma \gamma'\sigma}
  t^{\bf a}_{\gamma \gamma'} d_{{\bf i} \gamma \sigma}^{\dag}
  d_{{\bf i+a} \gamma' \sigma}
  -J_{\rm H} \sum_{\bf i}
  {\bf s}_{\bf i} \cdot {\bf S}_{\bf j} \nonumber \\
  &+& J_{\rm AF} \sum_{\langle {\bf i,j} \rangle}
  {\bf S}_{\bf i} \cdot {\bf S}_{\bf j}
  + \lambda \sum_{\bf i}
  (Q_{1{\bf i}}\rho_{\bf i} + Q_{2{\bf i}}\tau_{{\rm x}{\bf i}} 
  +Q_{3{\bf i}}\tau_{{\rm z}{\bf i}}) \nonumber \\
  &+& (1/2) \sum_{\bf i} (\beta Q_{1{\bf i}}^2
  +Q_{2{\bf i}}^2+Q_{3{\bf i}}^2),
\end{eqnarray}
where $d_{{\bf i}{\rm a}\sigma}$ ($d_{{\bf i}{\rm b}\sigma}$) is the
annihilation operator for an $e_{\rm g}$-electron with spin $\sigma$
in the $d_{x^2-y^2}$ ($d_{3z^2-r^2}$) orbital at site ${\bf i}$,
and ${\bf a}$ is the vector connecting nearest-neighbor (NN) sites.
The first term represents the NN hopping of $e_{\rm g}$ electrons 
with the amplitude $t^{\bf a}_{\gamma \gamma'}$ between
$\gamma$- and $\gamma'$-orbitals along the ${\bf a}$-direction
($t^{\bf x}_{\rm aa}$=$-\sqrt{3}t^{\bf x}_{\rm ab}$=
$-\sqrt{3}t^{\bf x}_{\rm ba}$=$3t^{\bf x}_{\rm bb}$=$1$ 
for ${\bf a}$=${\bf x}$ and
$t^{\bf y}_{\rm aa}$=$\sqrt{3}t^{\bf y}_{\rm ab}$=
$\sqrt{3}t^{\bf y}_{\rm ba}$=$3t^{\bf y}_{\rm bb}$=$1$
for ${\bf a}$=${\bf y}$, in $t^{\bf x}_{\rm aa}$ energy units).
In the second term, the Hund coupling $J_{\rm H}$($>$0) links
$e_{\rm g}$ electrons with spin
${\bf s}_{\bf i}=
\sum_{\gamma\alpha\beta}d^{\dag}_{{\bf i}\gamma\alpha}
\bbox{\sigma}_{\alpha\beta}d_{{\bf i}\gamma\beta}$
($\bbox{\sigma}$ are the Pauli matrices)
with the localized $t_{\rm 2g}$ spin ${\bf S}_{\bf i}$,
assumed classical and normalized to $|{\bf S}_{\bf i}|$=1.
The third term is the AFM coupling $J_{\rm AF}$
between NN $t_{\rm 2g}$ spins \cite{note:JAF}.
The fourth term couples $e_{\rm g}$ electrons
and MnO$_6$ octahedra distortions,
$\lambda$ is the dimensionless coupling constant,
$Q_{1{\bf i}}$ is the breathing-mode distortion,
$Q_{2{\bf i}}$ and $Q_{3{\bf i}}$ are, respectively, 
$(x^2$$-$$y^2)$- and $(3z^2$$-$$r^2)$-type JT-mode distortions,
$\rho_{\bf i}$=
$\sum_{\gamma,\sigma}
d_{{\bf i}\gamma\sigma}^{\dag}d_{{\bf i}\gamma\sigma}$,
$\tau_{{\rm x}{\bf i}}$=
$\sum_{\sigma}(d_{{\bf i}{\rm a}\sigma}^{\dag}d_{{\bf i}{\rm b}\sigma}
+d_{{\bf i}{\rm b}\sigma}^{\dag}d_{{\bf i}{\rm a}\sigma})$,
and
$\tau_{{\rm z}{\bf i}}$=
$\sum_{\sigma}(d_{{\bf i} a\sigma}^{\dag}d_{{\bf i}a\sigma}
-d_{{\bf i} b\sigma}^{\dag}d_{{\bf i}b\sigma})$.
%\cite{note:tau}.
The fifth term is the potential for distortions,
where $\beta$ is the ratio of spring constants for 
breathing- and JT-modes, treated here adiabatically
\cite{note:coulomb}.

%%%%%%%%%%%%%%%%%%%%%%%%%% FIG.1 %%%%%%%%%%%%%%%%%%%%%%%%%%%%%%%%%%%%%
\begin{figure}[t]
\centerline{\epsfxsize=7.truecm \epsfbox{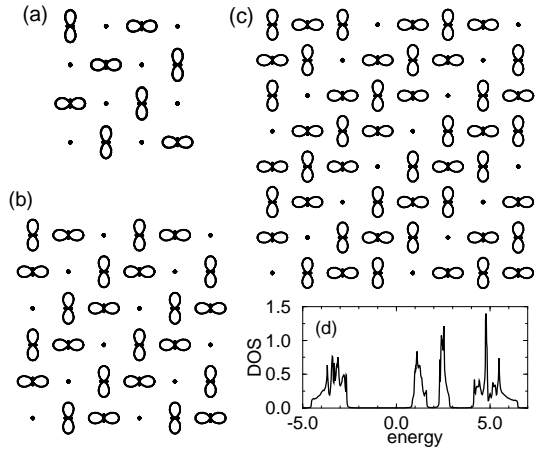} }
\label{fig1}
\smallskip
\caption{
Orbital arrangement in the CO FM phase at (a) $x$=1/2, (b) 1/3, and
(c) 1/4 with $\lambda$=2.0.
The charge density in the lower-energy orbital is shown,
with the size of the orbital being in proportion to this density.
(d) Total density of states vs energy for $x$=1/3 and $\lambda$=2.0.}
\end{figure}
%%%%%%%%%%%%%%%%%%%%%%%%%%%%%%%%%%%%%%%%%%%%%%%%%%%%%%%%%%%%%%%%%%%%%%%

First let us clarify the charge and orbital structure in the FM phase,
main issue of this paper.
If lattice distortions are not correlated in different sites, holes will
be simply distributed as uniformly as they can to lower the ground state
energy.
Namely, the stabilization of charge inhomogeneous structure such as 
stripes in the FM phase requires a proper treatment of
the {\it cooperative} JT effect.
A simple way to include such an effect is to optimize the displacement
of oxygen ions $u_{\bf i}^{\bf a}$ along the ${\bf a}$-axis
at site ${\bf i}$, instead of local distortions \cite{note:oxygen}.
Results at large $\lambda$ for the FM phase are shown in Fig.~1.
Using relaxation techniques to optimize $\{u\}$'s, 
{\it diagonal stripes} can be clearly
observed even in a spin FM regime, a surprising result.
The key ingredient to understand the presence of stripes in a spin
magnetized phase is the presence of the concomitant {\it staggered}
orbital order.
Individual holes doped into the $x$=0 FM orbital-ordered (OO)
state produce a distortion of the ordered background.
This energy lost is minimized if holes share
the distorted regions, forming complex patterns such as stripes.
A similar reasoning is usually applied to the rationalization of
stripe formation in nickelates and cuprates \cite{tranquada}, 
with the important conceptual difference that in those compounds
the background in which the stripes are created is spin AFM, i.e. 
the real spin is active, while in the present study the spin is FM and
the orbital background is active. If the orbital degree of freedom is
associated to a ``pseudo-spin'' up and down, then an analogy between
manganites and nickelates/cuprates can be established, replacing the
real spin of the latter by the orbital degree of freedom of the former.
In fact, with this analogy the pseudo-spin and charge structure
at $x$=1/4 (Fig.~1(c)) becomes the same as the real-spin and charge
stripe structure of hole-doped La$_2$NiO$_4$,
at the same hole density \cite{note:stripe}.
In the stripe phase, a gap of the order of the JT energy opens
around the Fermi level (see Fig.~1(d)).
Thus, ``pseudogap'' features 
may be expected as precursors for stripe formation
even at high temperatures such as $T_{\rm C}$.

Note that the stable charge-orbital stripes with the arrangement of
Fig.~1 can appear when the distance $d$ between diagonal hole arrays
is equal to $m a_0$, where $m$ is an integer and $a_0$ is the lattice
constant along the $a$-axis.
Since $d$=$a_0/x$ from Fig.~1, $x$ should be equal to 1/$m$ for the 
appearance of stable diagonal charge stripes.
For 1/($m$+1)$<$$x$$<$1/$m$, it seems possible (but at this early stage
it is a conjecture) that a mixed phase of two
charge-orbital ordered states with $x$=1/($m$+1) and 1/$m$ appears,
consistent with the phase separation scenario \cite{ps} and also
with recent synchrotron X-ray scattering measurements \cite{greven}.
Based on this scenario, the charge ordering observed in experiments
at $x$$\sim$0.3 in the FM regime may be understood as a mixture
of diagonal stripes at $x$=1/4 and 1/3 (Figs.~1(b) and (c)),
if those patterns are assumed to be stacked along the $z$-axis due to
the influence of $J_{\rm AF}$, a reasonable assumption based on previous
$x$=0.5 calculations\cite{hotta2}.
In the orbital correlation function $T^z({\bf q})$
=$\sum_{\bf i,j}e^{i{\bf q}\cdot({\bf i-j})}
\langle\tau_{z{\bf i}} \tau_{z{\bf j}} \rangle$,
peaks appear at ($\pi$$\pm$$\delta_m$, $\pi$$\pm$$\delta_m$)
with $\delta_m$=(1$-$2/$m$)$\pi$.
The deviation from ($\pi$,$\pi$) for $m$$>$2 is caused by a 
{\it $\pi$-shift} in the orbital order across the stripe (Fig.~1),
and it can be informally referred to as ``orbital incommensurability''
by analogy to the spin incommensurability found in cuprates and
nickelates.
It is important to remark that the present idealized
charge stripes will likely be
destabilized by thermal and/or quantum fluctuations.
Thus, in actual materials, it is expected that the stripes will be
{\it dynamical} as in cuprates and only vestiges of stripes
may be detected \cite{Belesi}, together with pseudogap features,
consistent with the phase separation
tendency for 1/($m$+1)$<$$x$$<$1/$m$.

Now let us consider the charge-orbital structure in the AFM phase
by using {\it non-cooperative} JT phonons. This will allow us to
report yet another striped state which could be observed experimentally,
this time with an overall zero magnetization.
As found at $x$=1/2 \cite{hotta2}, the AFM phase such as CE-type state
is not sensitive to the JT phonons treatment, since there exists a 
strong constraint due to the DE mechanism for the 
$e_{\rm g}$ electron kinetic motion,
masking differences in the character of JT phonons.
Note, however, that both the local lattice distortion and $t_{\rm 2g}$
spin direction should be determined independently at each site
by optimizing the total energy.
Using relaxation techniques to optimize $\{Q\}$'s and
$\{{\bf S}\}$'s at fixed electronic density \cite{note:non-coop},
the phase diagram at $x$=1/4 has been here obtained (Fig.~2(a)).
Its overall features can be understood from the competition between
the $e_{\rm g}$ electron kinetic energy and magnetic energy of
$t_{\rm 2g}$ spins regulated by $J_{\rm AF}$.
At small $J_{\rm AF}$ the system becomes FM to improve hole movement,
and a metal-insulator transition occurs at a critical value of $\lambda$,
separating FM CO and FM charge-disordered (CD) states.
In the other limit of large $J_{\rm AF}$, an AFM phase is stabilized
since the magnetic energy among $t_{\rm 2g}$ spins dominates.
The most interesting result of Fig.~2(a) is that at {\it intermediate} 
values of $J_{\rm AF}$, a novel spin-ordered state analogous to the
CE-type phase at $x$=1/2 has been found.
The complex optimized spin pattern is shown in Fig.~2(b).
A similar CE-like spin arrangement is also found at $x$=1/3 
(see Fig.~2(c)) \cite{note:zigzag1}.
These configurations are here called the ``zigzag'' AFM (Z-AFM) states,
since $t_{\rm 2g}$ spins form one-dimensional (1D) zigzag paths
where $e_{\rm g}$ kinetic energy is gained, stacking with
anti-parallel spins in the direction perpendicular to those
paths to gain magnetic energy.
Since this Z-AFM phase can take partial advantage of both energies, 
it is reasonable that it is stabilized in between the FM and AFM phases.

%%%%%%%%%%%%%%%%%%% FIG.2 %%%%%%%%%%%%%%%%%%%%%%%%%%%%%%%%%%%%%%%%
\begin{figure}[h]
\centerline{\epsfxsize=7.truecm \epsfbox{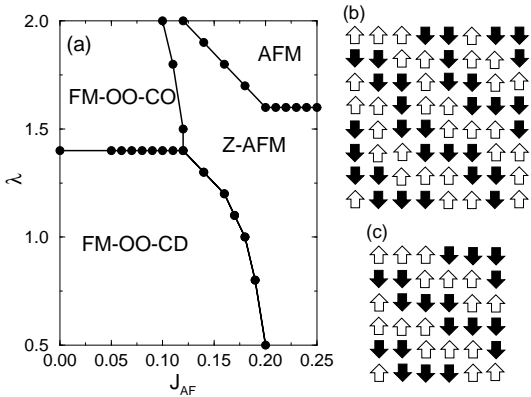} }
\label{fig2}
\smallskip
\caption{(a) Zero-temperature phase diagram at $x$=1/4 and large $J_{\rm H}$
(see text for notation).
The metallic FM-CD and insulating FM-CO boundary
is obtained monitoring the average JT distortions with increasing
$\lambda$.
(b) Ground-state spin pattern on an 8$\times$8 cluster for 
intermediate values of $J_{\rm AF}$ at $x$=1/4 (Z-AFM phase in (a)).
Open and solid arrows indicate up and down spins, respectively.
(c) Spin arrangement on the 6$\times$6 cluster
in the Z-AFM regime at $x$=1/3.
}
\end{figure}
%%%%%%%%%%%%%%%%%%%%%%%%%%%%%%%%%%%%%%%%%%%%%%%%%%%%%%%%%%%%%%%%%%

Consider now the charge and orbital structures of the phases in Fig.~2(a).
In the FM phase, for the values of $\lambda$ investigated, an OO phase
appears (not shown), irrespective of the charge structure.
As mentioned above, for non-cooperative JT phonons,
holes are distributed uniformly to lower the ground state energy,
compatible with the concomitant orbital order.
In the Z-AFM phase, the charge-orbital arrangements for large $\lambda$
are schematically shown in Figs.~3(a) and (b) for $x$=1/4 and 1/3,
respectively.
The local charge density is found to be constant along the diagonal
directions denoted by the broken lines in Fig.~3(a).
This result indicates a tendency toward the formation of
diagonal charge stripes in the Z-AFM phase for $x$=1/3 and 1/4.
In Fig.~3(c), the local charge densities on the diagonal lines, 
$n(\delta_1)$, $n(\delta_2)$, and $n(\delta_3)$, are shown
vs $\lambda$.
At small $\lambda$, diagonal charge stripes are observed,
concomitant with a peak around ${\bf q}$=($\pi$/2,$\pi$/2)
in the charge correlation function $n({\bf q})$ \cite{hill}.
On the other hand, at large $\lambda$, $n(\delta_3)$
is very small, while $n(\delta_1)$ and $n(\delta_2)$ are almost unity.
Namely, holes are mainly located along the line $\delta_3$, indicating
the formation of a clear diagonal charge stripe pattern
(see Fig.~3(a)), and the intensity of the $(\pi/2,\pi/2)$ peak
in $n({\bf q})$ becomes larger than at small $\lambda$.
Note that in the Z-AFM phase the $e_{\rm g}$ electron motion is
restricted to the 1D zigzag FM chains due to large $J_{\rm H}$.
When a finite electron-lattice coupling is introduced in this 1D system,
a Peierls instability should occur, leading to a charge-density-wave state.
Since in the Z-AFM phase the same chain is simply stacked along the diagonal
direction in the $x$-$y$ plane, diagonal charge stripes occurs naturally
\cite{note:CO}.

To understand the shape of the zigzag chain (Fig.2(b)), consider the limit
of $\lambda$=0 \cite{hotta2,hotta3}.
Even without phonons, by straightforward diagonalization it can be shown
that the zigzag chains have a spectra corresponding to
a {\it band-insulator} due to the periodic changes in hopping amplitudes
along zigzag chains ($t_{\rm ab}^{\bf x}$=$-t_{\rm ab}^{\bf y}$),
which induce gaps in the energy spectra \cite{hotta3}.
At $x$=1/4, there are nine independent possible types of zigzag chains
on an 8$\times$8 lattice with periodic boundary conditions.
Let us label these zigzag chains using ``bits'', 0 and 1,
representing the $x$- and $y$-directions, respectively.
The nine possible configurations are in Fig.~3(d).
The zigzag chain of Fig.~2(b) is given by the periodic sequence
``00101101'', and its energy is the lowest among the possible candidates,
since it provides the largest bandgap. 
When the electron-phonon interaction is adiabatically switched on, 
it is reasonable that the AFM phase Fig.~2(b) would still be stable
for a finite $\lambda$ range \cite{note:zigzag2}.
Note that in Fig.~3(d), excited states characterized by other zigzag paths
have small excitation energies, such as 0.001$\sim$0.01 in units 
of $t_{\rm aa}^{\bf x}$.
Thus, at temperatures as low as a few degrees Kelvin the ground-state
Fig.~2(b) can be easily distorted into other zigzag spin patterns.
This fragility of the ground-state may be related  with 
the ``spin glass'' features observed in single-layer experiments
\cite{moritomo}, leading to an overall orbital disordered state.
However, note that seven of the nine competing states, as well as the
Z-AFM phase, have diagonal stripes.
Then, even in a mixture of these states, the stripe direction is not random.
Thus, it is possible that indications of diagonal charge stripes may 
be present in the ``spin glass'' phase of single-layer manganites.

%%%%%%%%%%%%%%%%%%%%%% FIG.3 %%%%%%%%%%%%%%%%%%%%%%%%%%%%%%%%%%%%%%%%%%%%
\begin{figure}[t]
\centerline{\epsfxsize=7.truecm \epsfbox{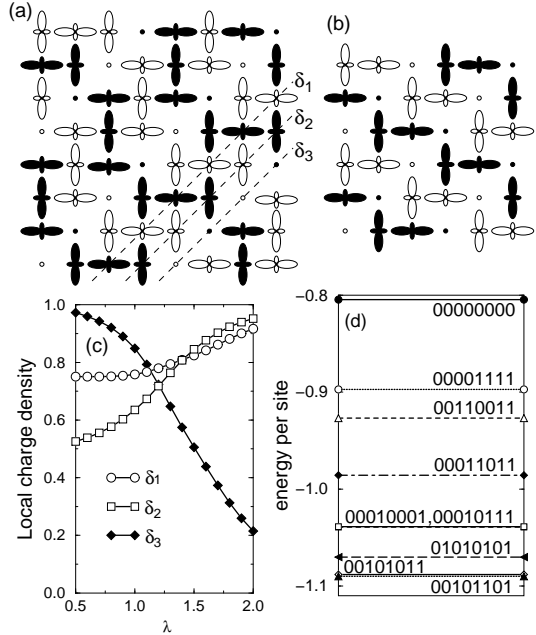} }
\smallskip
\label{fig3}
\caption{(a) Schematic view of spin-charge-orbital ordering in the
Z-AFM phase at large $\lambda$ and $x$=1/4.
Open and solid symbols indicate up and down 
$t_{\rm 2g}$ spins, respectively.
The lobes indicate $d_{3x^2-r^2}$- or $d_{3y^2-r^2}$-orbitals in the
Mn$^{3+}$ site, and the small circles denote the hole sites in which 
$d_{x^2-y^2}$-orbital is occupied by a tiny amount of charge.
(b) Same as (a), but for $x$=1/3.
(c) Charge densities (see (a)) vs $\lambda$ at $x$=1/4 in the Z-AFM,
where results at a given $\lambda$ are $J_{\rm AF}$ independent.
(d) Energies of several paths for $x$=1/4, $\lambda$=0, and 
$J_{\rm AF}$=0.25 in the Z-AFM regime (see text for notation).
}
\end{figure}
%%%%%%%%%%%%%%%%%%%%%%%%%%%%%%%%%%%%%%%%%%%%%%%%%%%%%%%%%%%%%%%%%%%%%%%

In summary, novel striped charge-orbital ordering has been found in
realistic models for manganites.
Diagonal stripes in the FM phase have been observed at densities $x$=$1/m$,
with $m$ an integer, and also in the CE-like Z-AFM phase.
The orbital degree of freedom orders in a $x$=0 staggered pattern in
between the stripes, playing a key role in stabilizing these structures,
similarly as the real spin does for stripes in nickelates and cuprates.
Our results have implications for the recently discovered short-range
charge ordering effects in neutron scattering experiments, as well as
for future experiments, particularly involving layered manganites.

The authors are grateful to P. Dai, M. Greven, J. Hill, and M. Kubota
for useful comments.
This work has been supported by the Ministry of Education,
Science, Sports, and Culture of Japan,
Fundaci\'on Antorchas, and grant NSF-DMR-9814350.

%%%%%%%%%%%%%%%%%%%%%%%% References %%%%%%%%%%%%%%%%%%%%%%%%%%%%%%%


\begin{references}

\bibitem{review}
See, e.g., {\it Colossal Magnetoresistive Oxides}, edited by Y. Tokura
(Gordon \& Breach, New York, 2000);
E. Dagotto {\it et al.}, to appear in Physics Reports.

\bibitem{ps}
S. Yunoki {\it et al.}, Phys. Rev. Lett. {\bf 81}, 5612 (1998);
A. Moreo {\it et al.}, Phys. Rev. Lett. {\bf 84}, 5568 (2000).

\bibitem{tunneling}
J. M. De Teresa {\it et al.}, Nature {\bf 386}, 256 (1997);
S. J. L. Billinge {\it et al.}, Phys. Rev. B{\bf 62}, 1203 (2000).

\bibitem{cubic}
S. Shimomura {\it et al.}, Phys. Rev. Lett. {\bf 83}, 4389 (1999);
P. Dai {\it et al.}, Phys. Rev. Lett. {\bf 85}, 2553 (2000);
C. P. Adams {\it et al.}, Phys. Rev. Lett. {\bf 85}, 3954 (2000).

\bibitem{note:vector}
The propagation vector is in the pseudo-cubic notation.

\bibitem{bi-layer}
L. Vasiliu-Doloc {\it et al.}, Phys. Rev. Lett. {\bf 83}, 4393(1999);
M. Kubota {\it et al.}, J. Phys. Soc. Jpn. {\bf 69}, 1986 (2000).

\bibitem{hotta1} T. Hotta {\it et al.},
Phys. Rev. B{\bf 60}, R15009 (1999).

\bibitem{hotta2} S. Yunoki {\it et al.},
Phys. Rev. Lett. {\bf 84}, 3417 (2000);
T. Hotta {\it et al.}, Phys. Rev. B{\bf 62}, 9432 (2000).

\bibitem{hotta3} T. Hotta, {\it et al.},
Phys. Rev. Lett. {\bf 84}, 2477 (2000).

\bibitem{kawano} S. Kawano {\it et al.}, 
J. Phys. (Paris) Colloq. {\bf 49}, C8-829 (1988).

\bibitem{murakami} 
B. J. Sternlieb {\it et al.}, Phys. Rev. Lett. {\bf 76}, 2169 (1996);
Y. Murakami {\it et al.}, Phys. Rev. Lett. {\bf 80}, 1932 (1998).

\bibitem{moritomo} Y. Moritomo {\it et al.},
Phys. Rev. B{\bf 51}, R3297 (1995).

\bibitem{note:JAF}
$J_{\rm AF}$ is small to make it compatible with experiments for the
fully hole-doped CaMnO$_3$ compound.
See also L. F. Feiner and A. M. Ole\'s, Phys. Rev. B{\bf 59}, 3295 (1999).

% \bibitem{note:tau}
% Note that the JT distortion with $E_{\rm g}$-symmetry does not couple with
% $\tau_{{\rm y}{\bf i}}$ belonging to the $A_{\rm 2u}$-representation.

\bibitem{note:coulomb} 
Coulomb interactions are effectively included in the large $\lambda$
regime (see T. Hotta {\it et al.} in Ref.~\cite{hotta2}).

\bibitem{note:oxygen}
Here only the bond stretching mode is assumed to occur and $\beta$ is
set as 2, a realistic value deduced from vibration energies for
breathing- and JT-mode phonons \cite{hotta1}.
Note that in an isolated 2D sheet there is no constraint for
$u^{\bf z}_{\bf i}$, the displacement of apical oxygens (AO),
but the real 2D system is embedded in a 3D environment,
suggesting that AO motion should be determined considering the
influence of other ions between sheets. In this work, for simplicity,
$u^{\bf z}_{\bf i}$ is set to zero, assuming that AO are fixed
in their positions by 3D effects.

\bibitem{tranquada} J. M. Tranquada {\it et al.},
Nature {\bf 375}, 561 (1995).

\bibitem{note:stripe}
To understand vertical/horizontal stripes, the cooperative analysis
should be carried out including modes for distortions other than the
bond-stretching mode. In the one-orbital model such stripes can also
be obtained (see Fig.III.d.6(b) in E. Dagotto {\it et al}. in
Ref.~\cite{review} and H. Aliaga {\it et al.}, cond-mat/0011342).

\bibitem{greven} S. Larochelle {\it et al.}, preprint; M. Greven,
private communication.

\bibitem{Belesi}
Once the stripes melt, together with the dynamical aspect of the charge,
the orbital order will no longer be long-ranged but short.
Then, melted stripes may lead to orbital liquid properties
(M. Belesi {\it et al.}, preprint).

\bibitem{note:non-coop}
To reduce the effort in our calculations, still capturing the main effect
of JT distortions, $\beta$ is set to $\infty$ effectively suppressing the
breathing mode, while $J_{\rm H}$=20 (but results at $J_{\rm H}$=$\infty$
are essentially the same).

\bibitem{note:zigzag1}
At $x$=1/3 and large $\lambda$, the zigzag
FM chain (Fig.~2(c)) is divided into small FM clusters
for non-cooperative JT phonons, while for cooperative case
the Z-AFM phase (Fig.~2(c)) is stable.
For large $\lambda$, the cooperative effect is crucial to stabilize
the Z-AFM structure, as well as to observe the
diagonal charge stripes in the FM phase.

% \bibitem{mizokawa}
% See also T. Mizokawa {\it et al.}, cond-mat/0011070.

\bibitem{hill}
Note that the charge density is nearly uniform at intermediate $\lambda$
such as 1.2 in Fig.~3(c). In this interesting regime, a peak
in the orbital correlations exist, while the analog in the charge sector
is not prominent. Such an OO but CD $x$=1/4 state may be related 
to those observed in recent experiments for Pr$_{1-x}$Ca$_x$MnO$_3$
(M. v. Zimmermann {\it et al.}, cond-mat/0007231).

\bibitem{note:CO}
In the Z-AFM phase, the charge pattern depends on $\lambda$ and
the shape of the FM path, but the orbital arrangement is mainly
determined by the zigzag geometry.

\bibitem{note:zigzag2}
Note that the results of Figs.~2(b,c) and the analysis of
competing zigzag paths have been obtained assuming either FM or
AFM links among localized spins, reasonable at large $\lambda$.
If the $t_{\rm 2g}$ spin directions are optimized for small $\lambda$,
spin patterns similar to Fig.~2(b) and (c) are stabilized,
but with spin directions slightly disordered.

\end{references}
\end{document}